\documentclass[nofootinbib, twocolumn, prd, superscriptaddress]{revtex4-2}
\usepackage{upgreek}
\usepackage{tikz}
\usepackage{dcolumn}
\usepackage{color,tcolorbox}
\usepackage{amsmath,mathrsfs,amssymb,esint}
\usepackage{dcolumn}
\usepackage{bm}
\usepackage{blindtext}
\usepackage{natbib}
\usepackage{graphicx}
\usepackage{rotating}
\usepackage{titlesec}
\usepackage{multirow}
\usepackage{mathtools}
\usepackage{soul}
\usepackage{amsmath}
\usepackage[pdfborder={0 0 0},colorlinks=true,linkcolor=blue,urlcolor=blue,citecolor=blue]{hyperref}
\colorlet{orange1}{green!10!orange!90!}
\usepackage{enumitem}

\def\be{\begin{equation}}
\def\ee{\end{equation}}
\def\bea{\begin{eqnarray}}
\def\eea{\end{eqnarray}}
\def\bi{\begin{itemize}}
\def\ei{\end{itemize}}

\def\mb{\mathbf}
\def\ba{\begin{equation}\begin{array}}
\def\ea{\end{array}\end{equation}}

\usepackage[T1]{fontenc}

\definecolor{ao(english)}{rgb}{0.0, 0.5, 0.0}
\usepackage{titlesec}
\usepackage{minted}
\usepackage{xcolor}
\definecolor{mintedbg}{rgb}{0.8,0.8,0.8}  
\begin{document}

\title{Dynamical Reconstruction of SPARC galactic halos within Self-interacting Fuzzy Dark Matter
%
%
%
}

\author{Milos Indjin}
\affiliation{
School of Mathematics, Statistics and Physics,
Newcastle University, Newcastle upon Tyne, NE1 7RU, United Kingdom}

\author{I-Kang Liu}
\affiliation{
School of Mathematics, Statistics and Physics,
Newcastle University, Newcastle upon Tyne, NE1 7RU, United Kingdom}

\author{Nick P. Proukakis}
\affiliation{
School of Mathematics, Statistics and Physics,
Newcastle University, Newcastle upon Tyne, NE1 7RU, United Kingdom}

\author{Gerasimos Rigopoulos}
\affiliation{
School of Mathematics, Statistics and Physics,
Newcastle University, Newcastle upon Tyne, NE1 7RU, United Kingdom}

\author{Aditya Verma}
\affiliation{
School of Mathematics, Statistics and Physics,
Newcastle University, Newcastle upon Tyne, NE1 7RU, United Kingdom}

\begin{abstract}
\noindent Fuzzy Dark Matter with non-zero quartic self-interaction (SFDM) is shown to be a viable model for simultaneously fitting 17 dark-matter-dominated galaxies from the SPARC database with a single $(m,g)$ point in the space of boson masses and self-coupling constants:
$\log_{10}\left(m \,[\mathrm{eV}/c^2] \right) = \log_{10}(1.98)-22^{+0.8}_{-0.6}$ and $\log_{10}\left(g \, [\mathrm{eV \, m}^3/kg] \right) = \log_{10}(9.08)-10^{+0.4}_{-1.2}$. 
This is based on the combination of an appropriately constructed static super-Gaussian profile for the inner galactic core (`soliton') region, and a Navarro-Frenk-White profile for the surrounding halo region. The explicit identification of a non-zero interaction strength may resolve issues of inconsistent constraints in non-interacting FDM. Our identification of these parameters enables the explicit {\em dynamical} reconstruction of potential host halos for such galaxies through numerical solution of the SFDM equations; we outline a {proof-of-principle  procedure via merger simulations} for two galaxies (UGCA444, UGC07866), and show that this yields viable rotation curves over 
a dynamical period of ${\cal O}(1) \, Gyr$.
\end{abstract}
\maketitle

\section{Introduction} 
Fuzzy Dark Matter (FDM) has emerged 
as an attractive dark matter model receiving increasing attention in the literature \cite{2016PhR...643....1M, Hui2021,2021A&ARv..29....7F,Hu2000,Schive2014,Schive2014a,Mocz2017,Marsh2015,Bernal2017,
Lin2018,Veltmaat2018,Levkov2018,
Mocz_2019,May2021StructureDynamics,Dome2023,
Matos_Review_2024,
OHare2024,
Chan2022,Yavetz2021,Liu2023,Bohmer2007,Chavanis2011,Chavanis2011Delfini,Harko2011zt,Rindler-Daller2014, Li2017, Desjacques:2017fmf,Dawoodbhoy2021,Shapiro2021,Glennon2020-2,Hartman2022,chanda_formation,marsh_formation,Chakrabarti2022,Indjin2024,
Amruth:2023xqj,
Mocz2023,Mocz2024, Moss:2024mkc, Chiu:2025vng}.
The FDM density profiles in the centres of halos include a core embedded in the surrounding halo~\cite{Schive2014, Schive2014a, Mocz2017} and bimodal profiles~(e.g.~\cite{Marsh2015,Bernal2017, Lin2018, Liu2023}) are typically used to describe them, commonly assuming no self-interactions.
Confrontation of FDM with observations can be done by various means~\cite{footnote_confront,
Painter:2024rnc,Elgamal_2024} 
with galactic rotation curves a common observable. In this latter context, a combination of the empirical profile~\cite{Schive2014} for the central core and the Navarro-Frenk-White ~(NFW) profile~\cite{Navarro1997} for the surrounding halo has been used and implemented to investigate the rotation curves~\cite{Bernal2017,Robles2018, Meinert2021,Khelashvili2023,Banares-Hernandez2023,Bar2018,Maleki2020,Bar2022,Rios2024} providing a credible alternative to the Burkert model of DM halos \cite{Burkert_1995}.

However, estimations of the FDM boson mass $m$ from observational rotation curves of galaxies are in tension with each other: studies of multiple galaxies find that a single value of the boson mass cannot explain the variety of curves from individual surveys, with different galaxies seemingly requiring a different boson mass in the range $10^{-23} - 10^{-20}$ eV/c$^2$ - see, \textit{e.g.}~\cite{Bernal2017,Meinert2021,Khelashvili2023} and references therein. Furthermore, the relation of the soliton core radius, $r_c$, to the core mass, $M_c$, in the form $r_c \sim 1/M_c$ predicted from FDM, seems at odds with observations \cite{Banares-Hernandez2023}. Therefore, if FDM is to be a viable explanation for the rotation curves, one must introduce some extra effect to cause a deviation from the $r_c - M_c$ scaling of $g=0$ FDM; one such possibility, which we will pursue here, is the inclusion of repulsive self-interactions.

Self-interacting FDM (SFDM, also known as Scalar Field Dark Matter, {or Self-Interacting Fuzzy Dark Matter (SIFDM)}, or BEC Dark Matter)~\cite{Bohmer2007,Chavanis2011,Chavanis2011Delfini,Indjin2024,Rindler-Daller2014, Li2017, Desjacques:2017fmf,Dawoodbhoy2021,Shapiro2021,Glennon2020-2,Hartman2022,Chakrabarti2022,Chavanis2022fvh,Harko2011zt,Mocz2023,Mocz2024,Painter:2024rnc,Moss:2024mkc,Stallovits:2024cpg,Indjin:2025thr,Lopez-Sanchez:2025atp}
is an extension of the original FDM model which allows for a local boson quartic self-interaction of strength $g \neq 0$. Focusing here on the SPARC database~\cite{SPARC}, we note that various authors~\cite{Bernal2017, Bar2018, Robles2018,Castellanos_2020,Harko_Rotation_2020,Chan2021, Bar2022, Street2022,Delgado2022,Dave_2023} have modelled its rotation curves via FDM.
Importantly, Ref.~\cite{Delgado2022} fitted a Gaussian profile to the inferred core region of the {dark matter component of the velocity profiles} from  a sample of 17 SPARC galaxies, systematically selected as the most dark matter dominated ones even in the central regions, i.e.~exhibiting no central baryonic bulge. This led to a \emph{single $(m,g)$ value} that could fit the rotation curves of all these galaxies within their respective inferred cores. Moreover, \cite{Dave_2023} found that, for a given boson mass, a range of self-interactions of comparable strength allow for a fit  to rotation curves of SPARC galaxies.

The aims of this paper are two-fold: 
Firstly, we extend the analysis of~\cite{Delgado2022}, fitting the rotation curves over their entire measured halo region. We use a bimodal ansatz consisting of a super-Gaussian (SG) profile 
in the inner region, transitioning naturally to a Navarro-Frenk-White (NFW) profile in the outer halo. 
The SG profile's parameters depend on an appropriate dimensionless interaction strength $\Gamma_g\ \sim gM_c^2m^3$ ~\cite{Indjin2024}, allowing for a better approximation of the impact of self-interactions on the soliton's peak density through the use of an exponent $\vartheta(\Gamma_g)$. 
Our analysis provides excellent static fits over the entire spatial extent of the rotation curves of all 17 dark-matter-dominated galaxies selected in \cite{Delgado2022}, with a {\em single} $(m,g)$ point identified (within error bars) as
\begin{eqnarray}
\log_{10}(m \,[\mathrm{eV}/c^2]) &=& \log_{10}(1.98)-22^{+0.8}_{-0.6} \nonumber 
\, ,\\
\log_{10}(g \,[\mathrm{eV} \, \mathrm{m}^3/\mathrm{kg}]) &=& \log_{10}(9.08)-10^{+0.4}_{-1.2} \nonumber \,, 
\end{eqnarray}
compatible with 
~\cite{Delgado2022,Dave_2023,footnote_Goswami}. Our results provide strong indication for the need to consider non-zero self-interactions while constraining the FDM boson mass.

Moreover, we demonstrate the existence of {\em dynamical quasi-equilibrium} SFDM solutions (created from mergers of smaller mass concentrations)
which are consistent with observed rotation curves, thereby outlining a {proof-of-principle procedure} for reconstructing a plausible host halo for a given (spherically averaged) rotation curve {from a merger simulation}. This is demonstrated for SPARC galaxies UGCA444, UGC07866 over a relevant timescale of ${\cal O}(1)$ billion years.

\begin{figure*}[t]
    \centering
    \includegraphics[width=0.7
    \linewidth,keepaspectratio]{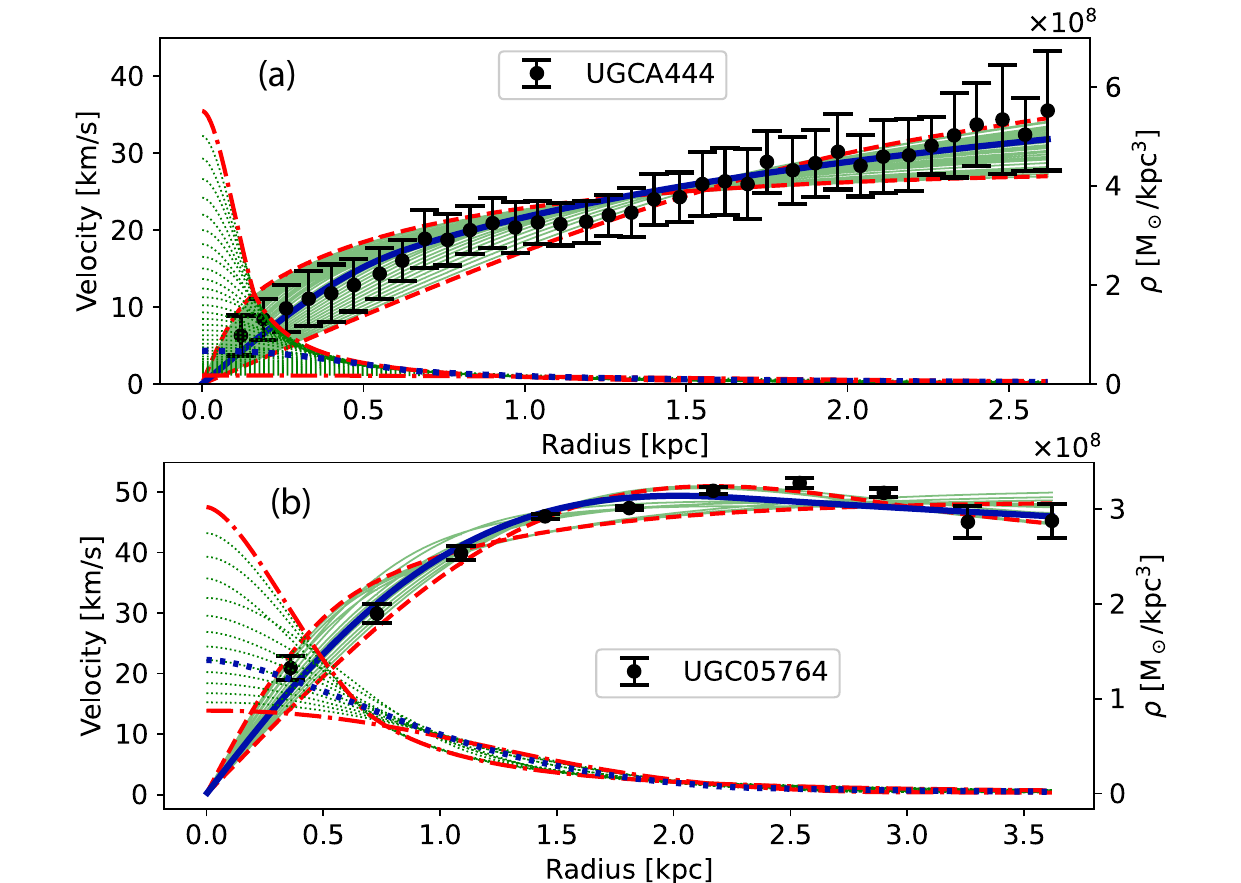}
    \caption{
    Rotation curves (left axis) and corresponding spatial profiles (right axis) for two different galaxies from the SPARC dataset~\cite{SPARC}: 
    optimal (minimum $\chi^2$) fits (blue) are highlighted from a broad list of acceptable fits (green), with the limiting red lines/points denoting incompatibility with our selection rules. 
    (Fitting details in 
    the Appendix.)
    \vspace{-0.4cm}
    } 
    \label{fig:Fig_1}
\end{figure*}

\section{Density and velocity profiles in SFDM}

\begin{figure*}[t]
    \centering   \includegraphics[width=0.7\linewidth,keepaspectratio]{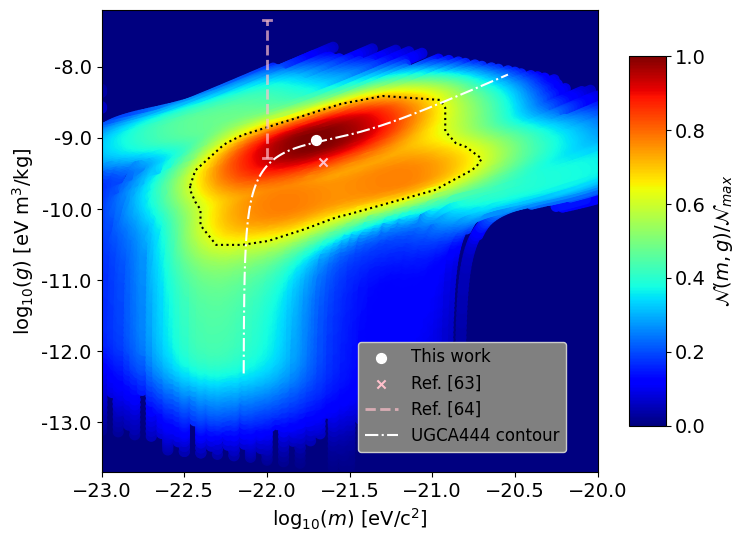}
    \caption{
    Density of acceptable soliton degeneracy contours   in $(m,g)$ space for the 17 probed dark-matter-dominated SPARC galaxies. 
    White circle: optimal value $(m_o,g_o)=(1.98\times10^{-22}\,\mathrm{eV}/c^2,\, 9.08\times10^{-10}\,{\rm eV \, m}^3/{\rm kg})$ around which most degeneracy contours   pass; 
    black dashed line: standard deviation estimate restricting parameter range for probed galaxies to $m \in (5 \times 10^{-23}, \, 1.25 \times 10^{-21}) \, {\rm eV}/c^2$, $g \in ( 5.6 \times 10^{-11}, \, 2.2 \times 10^{-9}) \, {\rm eV \, m}^3/{\rm kg} $. 
    Dot-dashed line depicts all acceptable points along a typical $(\rho_0,r_c)={\rm constant}$ degeneracy contour \eqref{eq:degeneracy} for a single galaxy. 
    We also show corresponding (compatible) $(m,g)$ estimate of Ref.~\cite{Delgado2022} ("$\times$"), and range of $g$ values of~\cite{Dave_2023} for a fiducial $m=10^{-22}$ ${\rm eV}/c^2$ (vertical dashed line).  
    }
    \label{fig:mg-heatmap}
\end{figure*}

Self interacting FDM is described by the mass density $\rho(\mb{r},t) = \left|\Psi(\mb{r},t)\right|^2$ of a complex scalar $\Psi(\mb{r},t)$ obeying the Gross-Pitaevskii-Poisson Equations (GPPE) \cite{Chavanis2011,Harko2011zt,Rindler-Daller2014}
\begin{eqnarray}
i\hbar\frac{\partial}{\partial t}\Psi(\mb{r},t) & = & \left[-\frac{\hbar^2\nabla_\mathbf{r}^2}{2m}+g\rho(\mb{r},t)+\Phi(\mb{r},t)\right]\Psi(\mb{r},t) \label{eq:GPE1} \\
\nabla_\mathbf{r}^2\Phi(\mb{r},t) & = & 4\pi Gm\left[
\rho(\mb{r},t)-\langle\rho(\mb{r},t)\rangle
\right] \;, \label{eq:Poisson_0}
\end{eqnarray}
 with $\Phi(\mb{r},t)$ the self-consistently determined gravitational potential.
Typical simulated DM halos via \eqref{eq:GPE1}-\eqref{eq:Poisson_0} consist of a quasi-static solitonic  core corresponding to the GPPE ground state~\cite{footnote_oscillations}, embedded in a larger dynamical halo which on average reproduces the NFW profile. 

The first step towards SFDM modelling of a given galaxy is to obtain an analytical equilibrium density profile, to act as the steady-state target of the SFDM simulation. 
We propose a bimodal, spherically symmetric, static profile to approximate the time-averaged density: 
$\rho(r) = \Theta(r_t-r)\rho_{SG}(r) + \Theta(r-r_t)\rho_{NFW}(r)$
where the central (solitonic) part of the SFDM profile is modeled by a super-Gaussian (SG) profile 
\begin{equation}\label{eq:superGauss}
    \rho_{SG}(r) = \rho_0(\Gamma_g) \exp\left[-\ln 2 \left( \frac{r}{r_c(\Gamma_g)}\right)^{\vartheta(\Gamma_g)} \right] \;,
\end{equation}
inspired by studies of self-bound Bose-gas droplets \cite{PhysRevA.103.053302}. This is analogous to the Einasto profile which has been used however to describe the whole halo in earlier literature \cite{Li:2020iib}. In our SG profile (see Appendix)  
the effect of boson self-interactions on the soliton shape is accounted for by making the peak density ($\rho_0$), the spatial extent ($r_c$) and the exponent ($\vartheta$) defining the slope of the profile implicit functions of an appropriately scaled interaction strength $\Gamma_g$ defined as~\cite{Indjin2024}
\begin{equation}
\Gamma_g = \frac{g}{g_*}\,\,,
\hspace{0.5cm} {\rm where} \hspace{0.5cm}
   g_* \equiv A_{0} \left(\frac{10 \hbar^4}{ G M_{c}^2 m^3}\right) \;. \label{eq:g_*}
\end{equation}
For the characteristic interaction strength $g_*$, the interaction energy of a spherical configuration equals its quantum kinetic energy~\cite{Chavanis2011,Indjin2024}:
for given ($M_{c},\, m$),
$\Gamma_g \sim 1$ marks the transition from weakly-interacting ($\Gamma_g \lesssim 1$) to strongly-interacting ($\Gamma_g \gg 1$) solitons.
$A_{0} = \sigma_0^2 / \nu_0 \zeta_0 \approx 20$ is fixed by the dimensionless prefactors (shape parameters) of the integrals defining the quantum pressure ($\sigma_0$), gravitational ($\nu_0$) and self-interaction energies ($\zeta_0$) in the absence of interactions -- see Ref.~\cite{Indjin2024}.
Here, the soliton mass, $M_c \approx M_{SG}$, where $M_{SG}$ the SG profile mass.
The well-known NFW profile, 
\be
\rho_{NFW}(r) = \rho_h \left[ (r/r_h) (1+r/r_h)^2 \right]^{-1} \;,
\label{eq:NFW}
\ee
takes over at a transition radius, $r_t$ (see, e.g.~\cite{Chan2022,Yavetz2021,Liu2023}). Physically relevant configurations require that at $r_t$: (i) profiles \eqref{eq:superGauss} and \eqref{eq:NFW} are equal, 
and (ii) the slope of the SG core is steeper than that of the NFW halo. 
The former constrains $\rho_h$; the latter ensures that the SG core is submerged below the NFW 
density at $r>r_t$, placing constraints on the admissible values of $r_t$, $r_c$ and $\vartheta$ in our fits.

The resultant profile is a \emph{model} that can be used in lieu of (time and angularly averaged) 
numerical solutions of \eqref{eq:GPE1}-\eqref{eq:Poisson_0}, involving a soliton core embedded within a $\rho \rightarrow r^{-3}$ halo~\cite{footnote_rh}. This static, spherically symmetric density profile leads to a rotation curve via
$v(r) = \sqrt{G M(r)\,/\,r}$, 
where $M(r)$ represents the mass contained within radius $r$. In particular, for the SG profile \eqref{eq:superGauss} we obtain
\be\label{eq:SGvel}
v_{SG}(r) = v_c \, \sqrt{ \frac{r_c}{r}\left[1-\frac{\Gamma\left(3 / \vartheta,\,\ln 2 \, (r/r_c)^\vartheta\right)}{\Gamma(3/\vartheta)}   \right]}
\ee
 where $\Gamma(a,x)$ is the upper incomplete Gamma function, such that $\Gamma(a)=\Gamma(a,0)$ is the Gamma function. 
The velocity of a test body orbiting at $r_c$ around a spherical object of mass $M_c$ equal to the soliton's mass becomes
\be
v_c = \sqrt{\frac{G M_c}{r_c}} = \sqrt{\frac{4 \pi G \rho_0 r_c^2}{\vartheta\left(\ln 2\right)^{3/\vartheta}} \, \Gamma\left(\frac{3}{\vartheta}\right) } \;.
\ee
For each $\vartheta$, \eqref{eq:SGvel} is a universal profile of the {\em inner parts} of a rotation curve,
extending the 
(Gaussian) limit~\cite{Chavanis2011, Delgado2022}.

\section{Dark Matter Dominated SPARC Rotation Curves with SFDM}\label{sec:III}
\begin{figure*}
    \centering
    \includegraphics[width=0.7\linewidth,keepaspectratio]{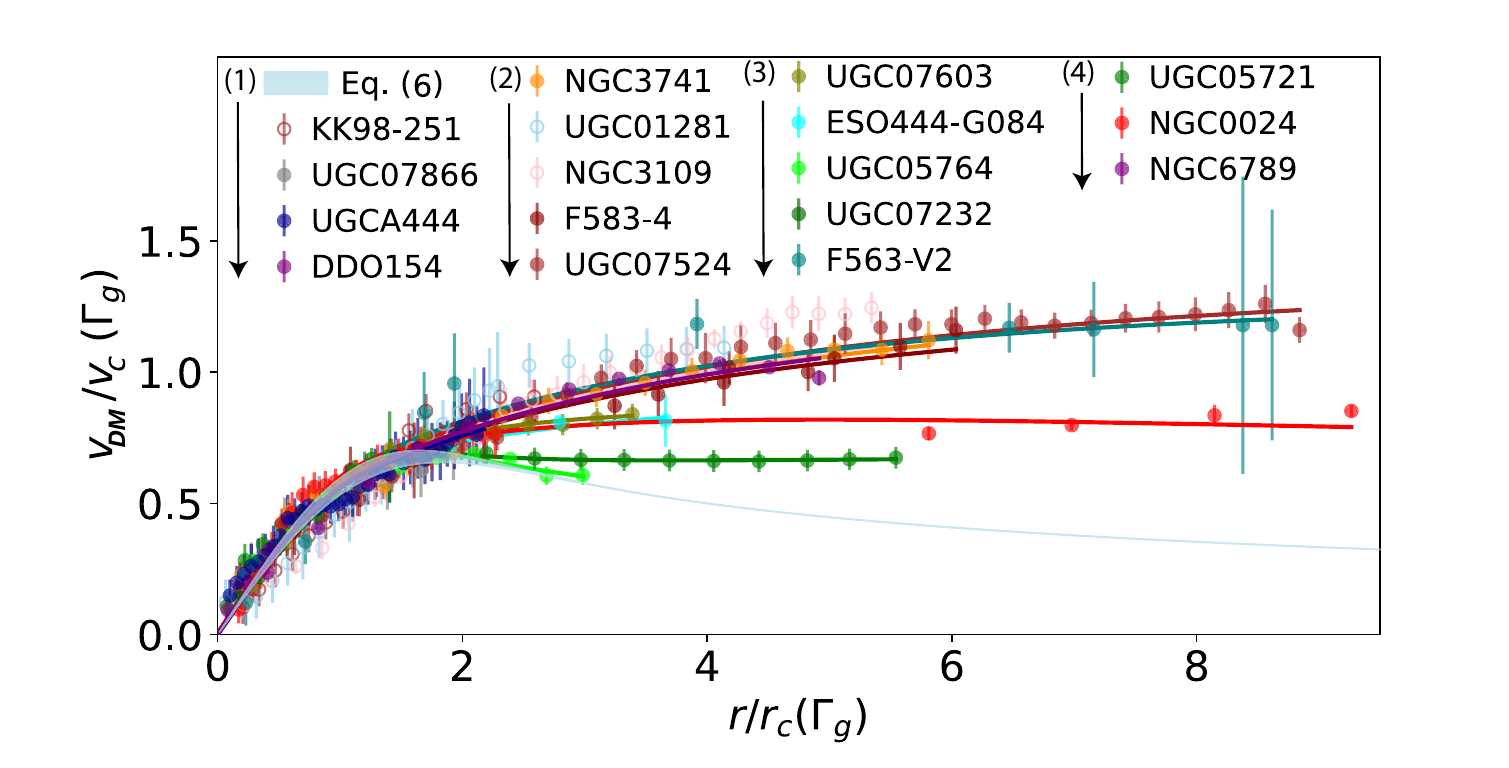}
    \caption{
    Optimal numerical dimensionless rotation curve fits (lines) to observational data (symbols) for all 17 SPARC galaxies for the optimal $(m_o,g_o)$ of Fig.~\ref{fig:mg-heatmap} {(i.e.~based on a 3-parameter fitting scheme)}. Galaxies are shown in order of increasing $\Gamma_g \in(4.8,1630.8)$. Three galaxies with a slight velocity over-/under-estimation in inner/outer core regions despite very good overall fits are shown by hollow symbols.
    Light blue band denotes universal prediction \eqref{eq:SGvel} based on the SG solitonic core density 
    valid for $r\lesssim r_t$ $\lesssim 2r_c$ here, 
    with band limits showing extrema of $\vartheta(\Gamma_g) = 1.86+0.68\tanh \left( \left[\log_{10}\Gamma_g - 0.6\right]/1.5\right) \in (1.97,2.26)$.
    Further details and individual galaxy rotation curve 3-parameter fits are shown in
    the Appendix.)
    }
    \label{fig:all_galaxies}
    \vspace{-0.4cm}
\end{figure*}

{Following~\cite{Delgado2022}, we consider a set of bulgeless SPARC galaxies with high quality rotation curves that were identified in that work as dark-matter dominated in the central region: For these galaxies  $\left\langle v_{\rm DM}/v_{\rm obs}\right\rangle  > 0.75$, where $\left\langle  \ldots \right \rangle$ refers to an average taken over the central data points of the rotation curves, as identified in~\cite{Delgado2022}, and $v_{\rm DM} = \sqrt{v_{\text{obs}}^2 - v_\text{g}|v_\text{g}| - \Upsilon_* v_* | v_*|}$ is the dark matter component of the rotation curves, with $v_\text{obs}, v_\text{g}$ and $v_*$ being the total observed, gas, and star contributions respectively, assuming a constant stellar mass-to-light ratio $\Upsilon_* = 0.47 M_{\odot}/L_{\odot}$   \cite{Delgado2022,McGaugh_2014}. In the present work we use all data points of the rotation curves, extending beyond the cores, finding that these galaxies are still classified as dark matter dominated according to the above criterion. Furthermore, practically all points of the rotation curves satisfy it individually with only a handful of data points (within the galactic cores) marginally dropping slightly below the $0.75$ threshold - see Fig.~\ref{fig:all_galaxies_separate}. 

We now apply our bimodal profiles to the arising 17 SPARC galaxies. 
If Eqs.~\eqref{eq:GPE1}-\eqref{eq:Poisson_0} are to be a fundamental description of dark matter, \emph{all galaxies} should point to a \emph{unique} $(m,g)$ value.
As a first step in identifying the plausibility of such a unique value, we initially allow $(m,g)$ (more specifically $m$ and $\Gamma_g$) to vary among individual galaxies (Sec.~\ref{sec:IV}).
While unphysical, such a step will allow us to identify those regions in $(m,g)$ space that can {\em potentially} fit all of our galaxies, as we describe below.
The observational viability of such a single $(m,g)=(m_o,g_o)$ value will then be tested (Sec.~\ref{sec:V}) by performing  physically relevant fits of the rotation curves based on such single $(m_o,g_o)$ value, aiming to get theoretical rotation curves acceptably close to those of all the galaxies in this sample.

}    

{Our initial heuristic 5-parameter search proceeds as follows:} We first search through the parameter space using the peak density  $\rho_0$ as our parent parameter and varying the remaining free parameters $\left\{\Gamma_g,m,r_t,r_h\right\}$. 
For a given galaxy, we first guess the initial peak density according to
$\rho_{0,\text{guess}} = (1/r_1^2) \partial/\partial r[r v^2/4 \pi G]$,
evaluated at the radial coordinate $r = r_1$ of the first data point. For this $\rho_0$ value, we begin a grid search across the remaining parameters in the order $\Gamma_g \rightarrow m \rightarrow r_t \rightarrow r_h$, varying 
in the ranges $\Gamma_g \in(0, 10^5)$ and $m \in (10^{-23},10^{-20})\,\mathrm{eV}/c^2$. $\Gamma_g$ 
fixes $\vartheta$, 
 while $\rho_0$, $\Gamma_g$, $m$ determine the soliton radius, 
 via
\begin{equation}
     r_c = \left[ B_{g} \frac{\hbar^2}{4\pi G \rho_0 m^2} \left(1+\sqrt{1+15\left(\frac{ A_0}{A_{g} }\right)\Gamma_g}  \right)   \right]^{1/4}  \label{eq:deg}
\end{equation}
where the self-consistently determined parameters $A_g$ and $B_g$ include the dependence of the shape parameters on interactions, with $A_{g} = \sigma_g^2/\nu_g \zeta_g \lesssim 1.3 A_0$, and $B_{g} \sim O(1)$ (details in~\cite{Indjin2024}, see also Appendix). 
This uniquely defines $\rho_{SG}(r)$ and $M_c$, with  $r_t$, $r_h$ defining the surrounding halo.

Acceptable sets of ($\Gamma_g$, $m$, $r_t$, $r_h$) are obtained for given $\rho_0$, via appropriate $\chi^2$ selection criteria yielding viable rotation curves consistent with observational data (details in the Appendix),
while $g$ is then obtained via~\eqref{eq:g_*}. Having obtained an acceptable set of parameter values for our initial $\rho_0$ choice, we repeat the procedure varying $\rho_0$. We thus obtain (for a given galaxy) a final range of plausible parameter values {defining velocity curves $v(r)$, corresponding to the profile $\rho(r) = \Theta(r_t-r)\rho_{SG}(r) + \Theta(r-r_t)\rho_{NFW}(r)$}, which are consistent with our selection criteria. The procedure is then repeated for all 17 galaxies considered here. 

{The above algorithm provides, for each galaxy in our sample, a set of rotation velocity curves, each corresponding to a specific value of the soliton mass $M_c$  (fixed by $\rho_0$, $\Gamma_g$ and $r_c$), the fundamental boson mass value $m$, and the fundamental self-coupling strength value $g$. Two such examples are given in Fig.~\ref{fig:Fig_1} alongside their corresponding density profiles. The resulting rotation curves identify \emph{individual scattered} points in $(m,g)$ space, each of which maps to a soliton that fits well one of the galactic cores in our sample. We now proceed to ask whether a fit to all galaxies with a global $(m_o,g_o)$ value, required within a physically consistent SFDM model, is possible. }

\section{Finding a global $(m_o,g_o)$}\label{sec:IV}

\begin{figure*}[t]
    \centering
    \includegraphics[width=0.7\linewidth]{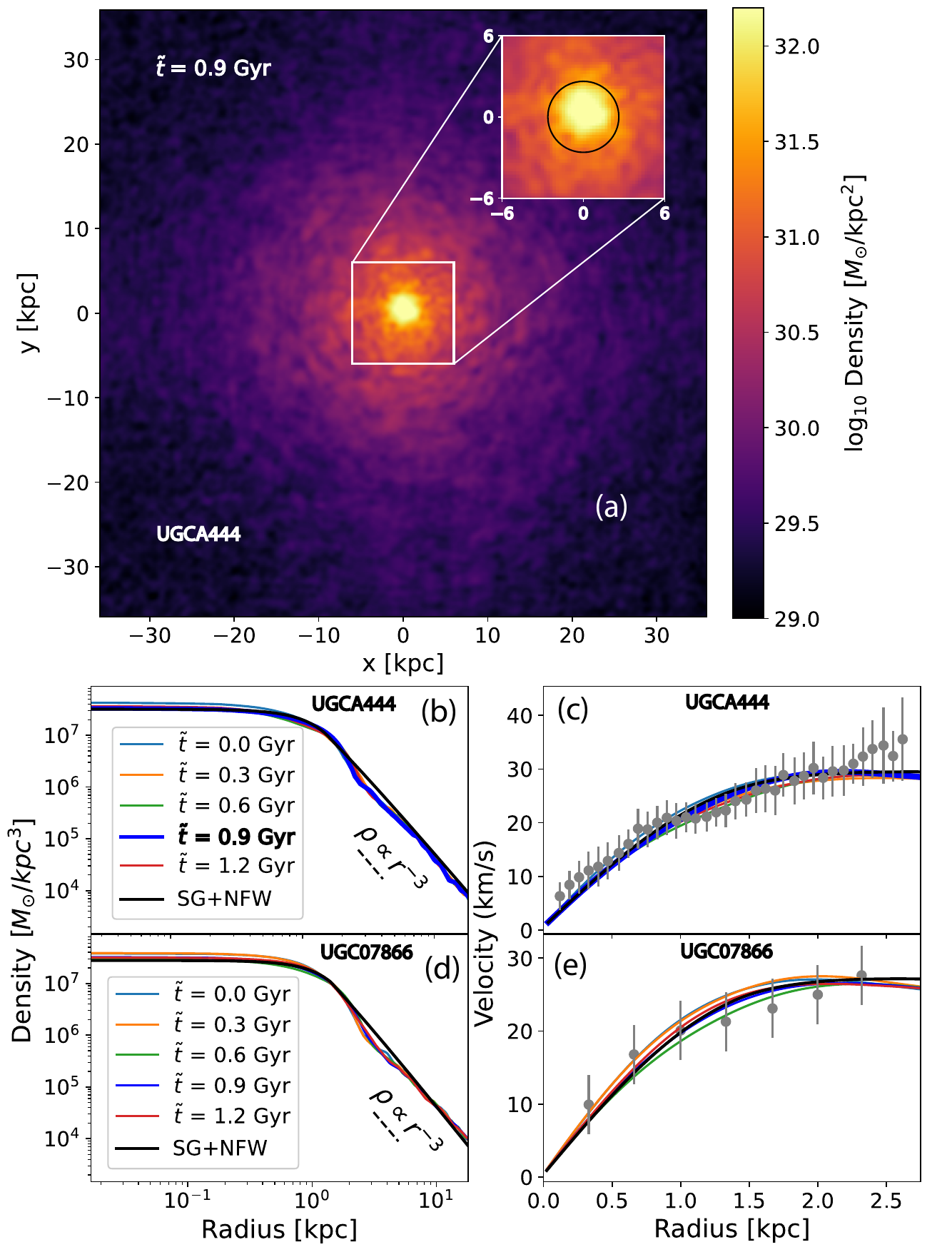}
    \caption{
    (a) Integrated density snapshot of UGCA444 projection corresponding to a viable rotation
    curve: black circle in inset highlights spatial extent over which observational velocity data exist. (b)-(c) Corresponding UGCA444 spatial profiles and rotation curves: shown are SFDM simulations at different sampling times $\tilde{t} \in (0, 1.2)$ Gyr (coloured lines), which are consistent with both observational data (points in (c)) and a static SG-NFW profile. (d)-(e) Corresponding plots for UGC07866. All plots constructed with the identified $(m_o,g_o)$.
    (GPPE simulations in a [$-34$,$+34$] ([$-24$,$+24$]) kpc box with $N = 320^3$ points and periodic boundary conditions~\cite{footnote_periodic}.)
    \vspace{-0.5cm}
    }
    \label{fig:ugca444_final}
\end{figure*}

{In searching for a plausible global $(m_o,g_o)$ value to fit all our galaxies, we first note that a soliton is mainly characterized by a given $\rho_0$ and $r_c$ which do not map to a unique point in $(m,g)$ space. More specifically, \cite{Indjin2024} showed that there exist \emph{degeneracy contours} in $(m,g)$ space, defined as loci of points which give rise to solitons with the same $M_c$, $\rho_0$ and $r_c$ (see also~\cite{Chavanis:2020rdo, Rindler-Daller2022}). They can be obtained via 
\be  
m(\Gamma_g) = \left[ B_{g} \frac{\hbar^2}{4\pi G \rho_0 r_c^4} \left(1+\sqrt{1+15\left(\frac{ A_0}{A_{g} }\right)\Gamma_g}  \right)   \right]^{1/2}  \label{eq:degeneracy}
\ee
given $\rho_0$, $r_c$ and replacing $\Gamma_g=\Gamma_g(g)$ by  \eqref{eq:g_*} where $M_c$ has been previously determined from the fits of Sec.~\ref{sec:III}.
One such degeneracy contour is shown in Fig.~\ref{fig:mg-heatmap} for one possible family of solitons with fixed $\rho_0$, $r_c$ and $M_c$ that could a-priori reside within UGCA444.  

Therefore, any $(m,g)$ point can be extended into such a degeneracy contour that traces a family of closely related solitons, all with mass $M_c$ and the \emph{same} $\rho_0,\, r_c$ parameters, but differing somewhat in their precise density profile due to variations of $\vartheta(\Gamma_g(g))$. Therefore, in order to obtain a complete coverage of the $(m,g)$ region giving rise to solitons that potentially fit the data, but which were not identified from our search in sec.~\ref{sec:III}, each individual $(m,g)$ point obtained there is now extended into its degeneracy contour. This allows us to cover all of the $(m,g)$ space region relevant to our rotation curves.}

{Although each previously obtained $(m,g)$ point has been extended to a degeneracy contour, these contours cannot be extended indefinitely. While the solitons associated to a degeneracy contour are very similar, they do have slightly different profiles which eventually impacts the resulting rotation curves. We thus traverse each degeneracy contour in discrete steps, recording the pairs $(m_j, g_j)$ (where $j$ defines the discretized steps in the $(m,g)$ parameter space), and verifying that the velocity (and corresponding density) profile generated by using $(m_j,g_j)$ and the
previously determined ($\rho_0, r_t, r_h$) values, also obey our selection criteria -- eliminating all points that do not. This leads to somewhat truncated degeneracy contours across which the computed rotation curves still fit well to the associated galaxy for a given $\rho_0, r_t$ and $r_h$. 
}
 
We can now use the degeneracy contours to identify the region of $(m,g)$ space which can support the largest number of fitted rotation curves from our sample. {Each discretized point $(m_j,g_j)$ from all of the degeneracy contours is converted to a smooth, localized density field using the 
\texttt{gaussian\_kde()}
function within the SciPy~\cite{SciPy} package, which turns these discretized points into Gaussian functions. We then assign a value at each point in $(m,g)$ space by the normalized sum of all surrounding contributing Gaussian functions. This procedure gives rise to a smooth `density function' $\mathcal{N}(m,g)$ which estimates the density of degeneracy contours passing through the neighbourhood of each $(m,g)$ point. The arising density map is shown in Fig.~\ref{fig:mg-heatmap} upon scaling the point density by its peak value, with the `highest-likelihood' region identified in a manner analogous to the $1\sigma$ contour of a Gaussian distribution: $\mathcal{N}(m,g)/\mathcal{N}_\text{max} > 0.605$.} 

{The function $\mathcal{N}(m,g)$ can be interpreted as follows: Each point in $(m,g)$ space a priori maps to a family of solitons of different total mass. Hence, since each degeneracy contour is associated with a given $M_c$, the number of degeneracy contours $\mathcal{N}(m,g)$ that pass from the neighbourhood of $(m,g)$ corresponds to the number of times a soliton of mass $M_c$ that can be mapped to $(m,g)$ has been found in our fitted rotation curves~\cite{footnote_probability}. Therefore the peak density value, $\mathcal{N}_\text{max}(m_o,g_o)$, indicates a heuristic estimate of an optimal value $(m_o,g_o)$
around which the highest density of degeneracy contours (i.e.~the highest number of identified solitons) exists. }
Interestingly, the maximum density point $(m_o, g_o)$ 
is indeed very close to the value quoted in~\cite{Delgado2022}.

\section{Rotation curves with $(m_o, g_o)$ \label{sec:V}}

{The procedure outlined in sections \ref{sec:III} and \ref{sec:IV} leads to an identification of a region in $(m,g)$ space that supports solitons which make a large number of appearances in our fitted rotation curves, with $\mathcal{N}(m_o,g_o)=\mathcal{N}_\text{max}$ mapping to solitons found most of the times in our sample. It should be stressed however, that rotation curves obtained by fitting with different values of $(m,g)$ for each individual galaxy cannot be physically acceptable since $m$ and $g$ are universal physical constants in the SFDM model. Hence, the previous fitted curves can only be considered as a heuristic route to a universal $(m, g)$. 

Having identified such a value in ($m_o$, $g_o$), we should now test its physical viability, in the sense of providing a good fit for {\em all} probed galaxies. To that effect we now proceed to the construction of fitted rotation curves for all galaxies using ($m_o$, $g_o$).
Specifically, we apply a variant of the algorithm described in section~\ref{sec:III} but now based on a fixed ($m_o$, $g_o$) (as expected for a self-consistent SFDM theory), and using $M_c$, instead of $\rho_0$, as our parent parameter. We therefore go through a 3-parameter grid search $M_c \rightarrow r_t \rightarrow r_h$ (as opposed to our initial heuristic 5-parameter search) using the same acceptance criteria as before. }

{The resulting best fit galaxy rotation curves with ($m_o$, $g_o$) can be seen in Fig.~\ref{fig:all_galaxies} where all 17 galaxies are simultaneously shown, after rescaling $v$ and $r$ by $v_0$ and $r_c$ respectively for each galaxy. The common core to all galaxies is clearly visible while the outer rotation curves, individual to each galaxy through $r_t$ and $r_h$, are also fit well. To estimate the goodness-of-fit we compute the Relative Root Mean Square Error (RRMSE) of the fitted rotation curves ${\rm D}[v(r)]$, where 
\begin{equation}\label{eq:rrmse}
    {\rm D}[v] = \sqrt{\frac{1}{n} \sum_i \frac{\left( v(r_i) - v_{{\rm DM},i} \right)^2}{v_{{\rm DM},i}^2} } \,,
\end{equation}
$v(r)$ is the numerically evaluated rotation velocity associated to our fitted density profiles $\rho(r)$ and $v_{\rm DM}$ is the dark matter component of the observed rotation curves. For our galaxies, we find for the best fit rotation curves ${\rm D} [v_{\rm bf}]\simeq 0.13\pm 0.05$. More details and the fit curves for individual galaxies can be found in the appendix.}

\section{{Host Halos for SPARC Galaxies UGCA444 $\&$ UGC07866 from mergers}} 
The above approximate analytical SG-NFW profiles fitting the velocity curves can be interpreted as approximations to numerical solutions of \eqref{eq:GPE1}-\eqref{eq:Poisson_0}.
We demonstrate this by constructing numerical solutions compatible with the rotation curves of two of the lower mass galaxies (UGCA444, UGC07866). Our procedure is based on first identifying `target' density profiles through SG-NFW fits, 
and then generating those through dynamical GPPE {merger} simulations. Below we outline the key elements, with technical details given in the Appendix. As far as we are aware, reconstruction of known galaxies attempting to replicate observational signatures has only been done in~\cite{Rios2024, Marsh2025} but by the stacking of eigenmodes and not by fully solving \eqref{eq:GPE1} and \eqref{eq:Poisson_0} as done here. See also Refs.~\cite{Lin2018, Yavetz2021,Yang2024} which are not however directly related to observed rotation curves.

The main challenge {in reconstructing a target galactic halo from such merger GPPE simulations} is that we do not know the total halo mass, or kinetic energy -- despite knowing (from SG-NFW fits) the  target density profile, and thus the potential, self-interaction and (soliton) quantum kinetic energy densities. Absence of consensus on the core-halo relation for cosmologically relevant halos~\cite{Schive2014a, Mocz2017, Nori2020, Chan2022, Mina2022, Zagorac2023, Liao:2024zkj,Blum:2025aaa},
implies the halo mass of the galaxy in our numerical simulations is necessarily dependent on the spatial extent of our numerical box. Boundary effects are minimised by imposing a projected density of the SG-NFW profile at the box edge  
$\lesssim \mathcal{O}(10^{-4})\, \rho_0$, 
placing constraints on how massive a halo we can simulate; nonetheless, such constraints can be minimized by the freedom provided by the (often sizeable) observational velocity error bars, which allows us to pick for our SG-NFW target fit those with the smallest value of $r_h$ (i.e.~the smallest mass in the region beyond observational data).

Lack of knowledge of the halo's total (conserved) energy (but see \cite{Blum:2025aaa}) can be bypassed by noting that as the soliton constitutes only a small fraction of the total mass ($M_{c}/M_{\text halo} \ll 1$), the halos are highly-excited, or "puffy", implying ${\left| E \right|}/{\left| E_0\right| }\ll 1 $,
with $E<0$ the energy of the whole halo and $E_0$  the ground state energy of the system (\textit{i.e.}~of a single isolated soliton with the same total mass). 
This guides the initial condition of our GPPE dynamical simulations: 
firstly we partition
the total mass
into $\mathcal{O}(10)$ equal Gaussian-shaped mass lumps, randomly distributed within a subset of the box, but in a manner which ensures their centre of mass is located at the centre of the box.
The configuration energy is controlled via the position/number of lumps (fixing gravitational potential energy), while changing the size of each lump modulates the quantum kinetic and self interaction energies. 
Simulations with an appropriate variety of initial conditions (allowing for slight variations in total mass and energy) are propagated dynamically for an extended time of $ {\cal O} (10) \, \text{Gyr}$, to ensure that the quasi-equilibrium core-halo structure arising from the gravitational coalescence of the initial masses~\cite{Liu2023} leads to the desired parameters: we find ${\left| E \right|}/{\left| E_0\right| }\sim \mathcal{O}(10^{-2})$ to be sufficient for viable rotation curves.
To visualize UGCA444 (UGC07866) we use the SG-NFW target profiles with: $\rho_0 = 3.20 \,(2.79) \times 10^{7} \, M_\odot / \text{kpc}^3$, $r_c = 1.2\, (1.2)\, \text{kpc}$, $r_t = 1.50 \, (1.42) \, \text{kpc}$, $M_c = 0.51 \, (0.45) \times 10^{9} M_\odot$;
with $M_\text{\rm box} = 2.15 \, (1.85) \times 10^{9} M_\odot$.

Once a quasi-equilibrium state of a galaxy similar to our target galaxy is reached,
we begin sampling the shell-averaged density profiles in time, comparing the computed velocity curves to the observational data: we identify a temporal regime of ${\cal O}(1)$ Gyr (larger than observationally relevant ones~\cite{Levkov2018, Eggemeier2019, Chen2021, Rios2024}) over which the rotation curves remain  consistent with observational data (within error bars), with densities recovering the $r^{-3}$ section of the NFW profile at large $r$, see Fig.~\ref{fig:ugca444_final}.
The projected density plot (integrated along $z$)
shows a typical generated density snapshot for UGCA444, highlighting (black circle in inset) what a small volume of the total reconstructed dark halo the observational data probes. The final observational data point is located outside the soliton core, at $r \sim 2r_c$.

\section{Conclusions}
We showed that a bimodal SG-NFW profile fit to the rotation curves of the 17 most relevant dark matter dominated galaxies from the SPARC database allows us to hone in on a single $(m_o,g_o)$ pair which reproduces core-halo density profiles {reasonably consistent} with their rotation curves. This resolves a stated inability of {\em non-interacting} FDM to fit {\em multiple} rotation curves with a {\em single} boson mass~\cite{Khelashvili2023}. Our optimal $(m_o,g_o)$ values are compatible with those in~\cite{Delgado2022, Dave_2023}, offering positive evidence towards the need for including non-zero self-interactions in FDM. 

We also proposed a scheme for {\em on demand} dynamical reconstruction of observationally-relevant cored galactic halos within fully dynamical simulations of self-interacting FDM (SFDM), demonstrating this at proof-of-principle level  by reproducing quasi-equilibrium rotation curves for galaxies UGCA444 and UGC07866 over observationally-relevant timescales of  $\mathcal{O}(1)$ Gyr. 
This allows going beyond spherically averaged, approximate profiles and may open avenues for assessing the impact of the host halo's {\em fully dynamical} FDM field $\Psi(\mathbf{r},t)$ on the baryonic matter making up the observable galaxy. {Of course, our halos are products of heuristic merger simulations. Assessing whether such numerical solutions can also arise from realistic cosmological initial conditions and in an expanding universe (see, e.g.~\cite{Mina2022,Nori2020,Liao:2024zkj} for the $g=0$ case), is something that will require further work. However, our study can be considered as evidence that such solutions may indeed be relevant for describing dark matter halos. }

The small sample examined here was chosen among the SPARC galaxies so as to have little baryonic matter in their centres. Our results for $(m_o,g_o)$ should be scrutinized with a larger sample of similar galaxies and more sophisticated statistical methods, {which will be the subject of future work}. Nevertheless, we believe that they provide an indication that self-interactions may be important in dark matter dynamics even beyond rotation curves -- for example, they introduce degeneracies in the dynamics of linearized density perturbations \cite{Proukakis:2023txk} -- and hence that there is need for a more thorough examination of self-interactions in other contexts where FDM has been studied, e.g.~constraining $m$ from other larger scale/higher $z$ cosmological observables \cite{Rogers:2020ltq}. 

{The restriction to only considering galaxies with low central baryon content imposed in our work was aimed at avoiding influences of the latter as much as possible, either via feedback mechanisms or direct interaction with the dark matter.} It would be interesting to examine how such self-interacting halos might relate to the phenomenological universal galactic rotation curves for broader galaxy types and observed scaling relations \cite{10.1093/mnras/stw3055,Salucci:2018hqu,Burkert:2020laq, Almeida:2025pwb}. Interestingly, we note that the values of the product $\rho_0 r_c$ in our SFDM cores are compatible with those recently reported in \cite{Almeida:2025pwb}.

\section*{Acknowledgements}
Funding was provided by the Leverhulme Trust Grant RPG-2021-010 (NPP, GR), and STFC 
grants ST/W001020/1 (IKL) and ST/W006790/1 (AV). We thank Thomas Bland for the suggestion of a Super-Gaussian profile and Markus Rau, Dani Leonard, Marika Asgari, Giorgos Vasdekis, and Ramit Goolry for discussions.

\section*{Data Availability}
Data supporting this publication can be openly accessed under an `Open Data Commons Open Database License' [...to appear on publication...]

\appendix

\titleformat{\section}
  {\normalfont\small\bfseries}
  {Appendix \Alph{section}:}
  {1em}{}

\renewcommand{\thesubsection}{\Alph{section}\arabic{subsection}}

\makeatletter
  \renewcommand{\p@subsection}{}
  \renewcommand{\theHsubsection}{\Alph{section}\arabic{subsection}}
\makeatother
\section{Additional Characterization Details}

Here we provide further details of our Super-Gaussian (SG) profiles (Sec.~\ref{app: SG Prof. Char.}), and bimodal SG-NFW fitting characterization with particular attention paid to the selection criteria used in our fitting (Sec.~\ref{app: SG-NFW Fits and Selection Criteria}), and the role of $r_h$ in constraining the rotation curves in the outer halo (Sec.~\ref{app: Impact of r_h term}), thus motivating the choice of parameters in our dynamical SFDM simulations. We also give an example of our initial state preparation and further details on the dynamical procedure of our GPPE simulations (Sec.~\ref{app: Dynamical GPPE Sim.}).
Finally, for clarity, we give details of all extracted values for each of the 17 dark-matter-dominated SPARC galaxies, showcasing for our identified $(m_o,g_o)$ values the optimal rotation curve fits for each galaxy individually (Sec.~\ref{app:fits}).

\begin{figure}[t!]
    \centering
    \includegraphics[width=1\linewidth]{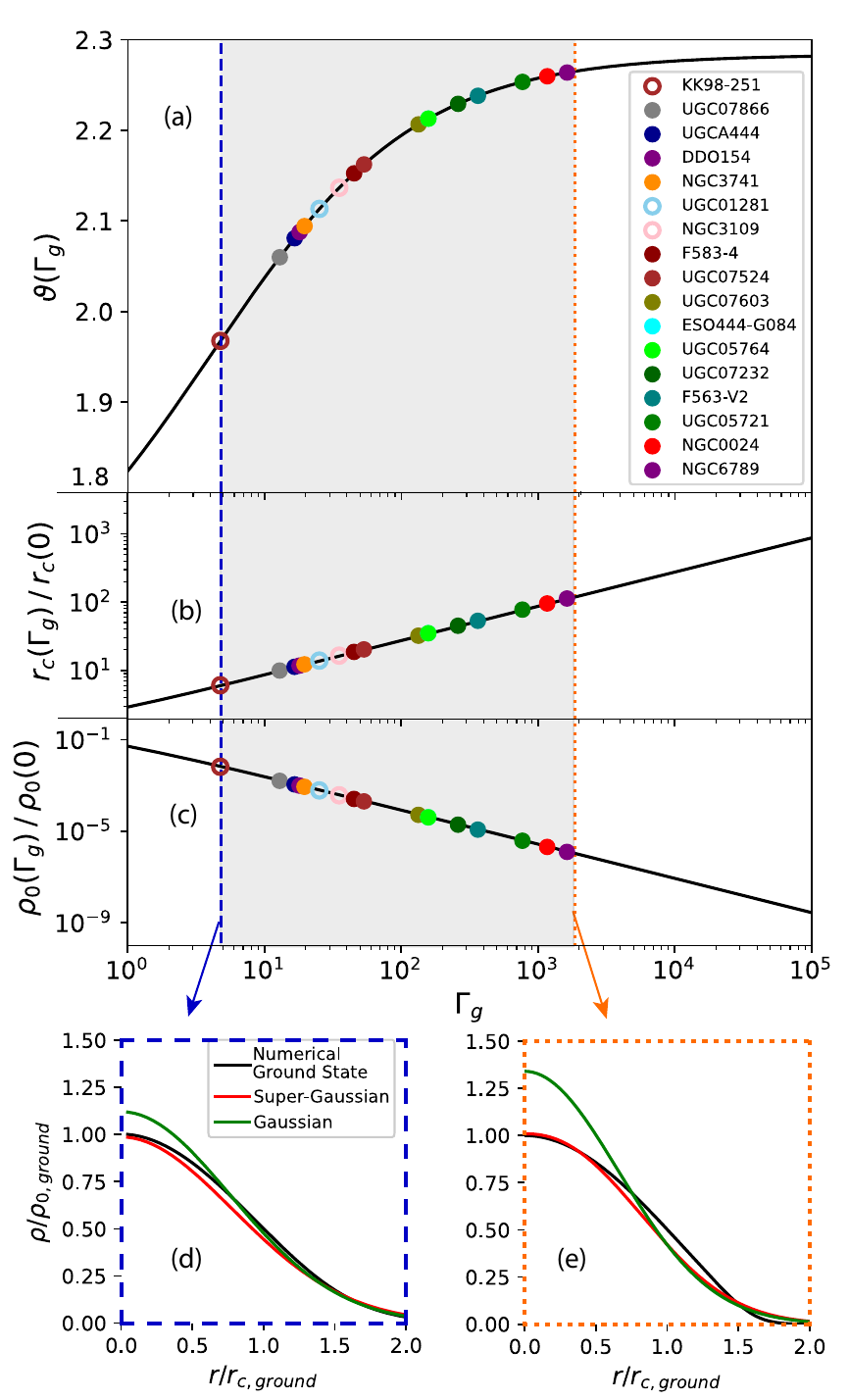}
    \caption{
    Analysis of the SG profile [main text, Eq.~(3)]:
    (a) Exponent $\vartheta$,
    (b) scaled core radius, and (c) scaled peak density 
    as a function of dimensionless interaction strength $\Gamma_g$:    
    inferred values for the 17 SPARC galaxies (symbols) are shown, with all galaxies lying within the grey band for which $\Gamma_g > 1$ and $\vartheta \gtrsim 1.97$.
    (d)-(e) Comparison of SG (red) and Gaussian (green) spatial profiles to renormalized GPPE numerical ground state solutions \cite{Indjin2024} encompassing the probed SPARC dataset. }
    \label{fig:Y}
\end{figure}

\subsection{Super-Gaussian Profile Characterization}
\label{app: SG Prof. Char.}
Our previously-studied~\cite{Indjin2024} extended data simulation set 
of true ground state solutions of the GPPE equations [Eqs.~(1)-(2) in the main text] with $\Gamma_g \in (0,\, 10^5)$,  can be conveniently re-analyzed by a SG profile [Eq.~(3) in main text].
For a fixed $M$, $m$ and $g$ (which fixes $\Gamma_g$), we self-consistently obtain the value of $\vartheta(\Gamma_g)$ matching the numerical profile.
This
yields the following empirical relationship for the SG exponent, 
\begin{equation}
    \vartheta(\Gamma_g) = \frac{\vartheta_0+\vartheta_{TF}}{2}+\frac{\vartheta_{TF}-\vartheta_{0}}{2}\tanh\left( \frac{\log_{10}\Gamma_g-a}{b}\right) \;, \label{eq:upsilon}
\end{equation}
with $a=0.6$, $b=1.5$ (with $\vartheta_0 = 1.62 $, $\vartheta_{TF} = 2.3 $).
Such dependence is shown in Fig.~\ref{fig:Y}(a)  by the black line, restricted here to the relevant `interacting' regime $\Gamma_g \ge 1$.
The extracted dimensionless interaction values for the SPARC galaxies shown in Fig.~3 of the main text, lie in the range $\Gamma_g \in (4.8,\, 1630.8) \gg 1$, with a corresponding SG profile exponent $\vartheta \in [1.97, 2.26]$ (subsequent Table~\ref{table:parameters}).

 The analytical expressions governing the dependence of the spatial extent $r_c$ and peak density $\rho_0$ on $\Gamma_g$ discussed in Ref.~\cite{Indjin2024} can  be recast in the form
 \begin{equation}
\frac{r_c(\Gamma)}{r_c(0)} = \left (\frac{B_g}{B_0} \right)
\left( \frac{\eta_g}{\eta_0} \right)
\left[ \frac{1}{2} \, \left(1+\sqrt{1+15\,\left( \frac{A_0}{A_g} \right) \Gamma_g}\right) \right] \;, \label{eq:radius_g-ratio}
\end{equation}
\begin{equation}
\frac{\rho_0(\Gamma_g)}{\rho_0(0)} = 
\left( \frac{\eta_g}{\eta_0} \right)^{-1}
\left[\frac{r_c(\Gamma_g)}{r_c(0)} \right]^{-3} \;. \label{eq:rho_g-ratio}
\end{equation}
The combined prefactors of the square brackets appearing in the above equations are ${\cal O}(1)$; specifically, for the SG profile, their ranges are: $(1,\,1.6)$ in Eq.~\eqref{eq:radius_g-ratio}, and $(1,\,1.8)$ in Eq.~\eqref{eq:rho_g-ratio}.
The non-interacting spatial extent and peak density appearing above are given by
\begin{eqnarray}
    r_c(0) = \frac{2\sigma_0}{\nu_0}\frac{\hbar^2}{G M m^2} \,\, , \hspace{0.5cm}
    \rho_0(0) = \frac{M_c}{4\pi\eta_0 r_c(0)^3},
\end{eqnarray}
and the quantities $A_g = \sigma_g^2 / \nu_g \zeta_g$, and $B_g = \sigma_g / \nu_g \eta_g $ introduced in the main text
satisfy the respective ranges
$A_g  \in (21.3, 27.0)$, and $B_g \in (1.1, 3.2)$ (while, in the absence of interactions, $A_0 = 21.3$).
For completeness, we remind the reader here 
that the shape parameters $\eta_g$, $\sigma_g$, $\nu_g$, $\zeta_g$ appearing in these expressions have already been defined in Ref.~\cite{Indjin2024} through their relevant energy integrals as
\begin{align}
    M_c &= 4\pi \int_0^\infty \rho r^2 \; dr = \eta_g \, \left( 4 \pi \rho_0 r_c^3 \right) \, ,
\end{align}
\begin{align}
    \Theta_Q &= \frac{2\pi \hbar^2}{2m} \int_0^\infty \left|\frac{\partial}{\partial r} \sqrt{\rho}\right|^2 \; dr = \sigma_g \left( \frac{\hbar^2 M_c}{m^2 r_c^2} \right) \, ,
\end{align}
\begin{align}
    W &= - \frac{8\pi G}{2} \int_0^\infty r M_c(r) \rho \; dr = -\nu_g \, \left(\frac{G M_c^2}{r_c} \right) \, ,
\end{align}
\begin{eqnarray}
    U = \frac{2\pi g}{m} \int_0^\infty \rho^2 r^2 \; dr  
    = \zeta_g \, \left( \frac{M_c^2 g}{2 m r_c^3} \right) \, .
\end{eqnarray}
Note that the exact value of such (interaction-dependent) shape parameters ($\eta_g$, $\sigma_g$, $\nu_g$, $\zeta_g$) depends on the chosen density profile $\rho(r)$.
For the specific SG profile discussed here, such shape parameters  become
implicit functions of the SG exponent $\vartheta(\Gamma_g)$, e.g. $\eta_g = \eta(\vartheta(\Gamma_g))$.


The arising SG profiles provide an accurate representation of the numerical SFDM solutions  of an isolated solitonic core in the presence of non-zero interactions: this can be clearly seen in Fig.~\ref{fig:Y}(d)-(e), where the SG profiles (red lines) match the SFDM solitonic cores (black lines), demonstrating a notable deviation in the central region and peak density from the Gaussian profiles (green lines) with increasing interactions.

\begin{figure}[t]
    \centering
    \includegraphics[width=\linewidth,keepaspectratio]{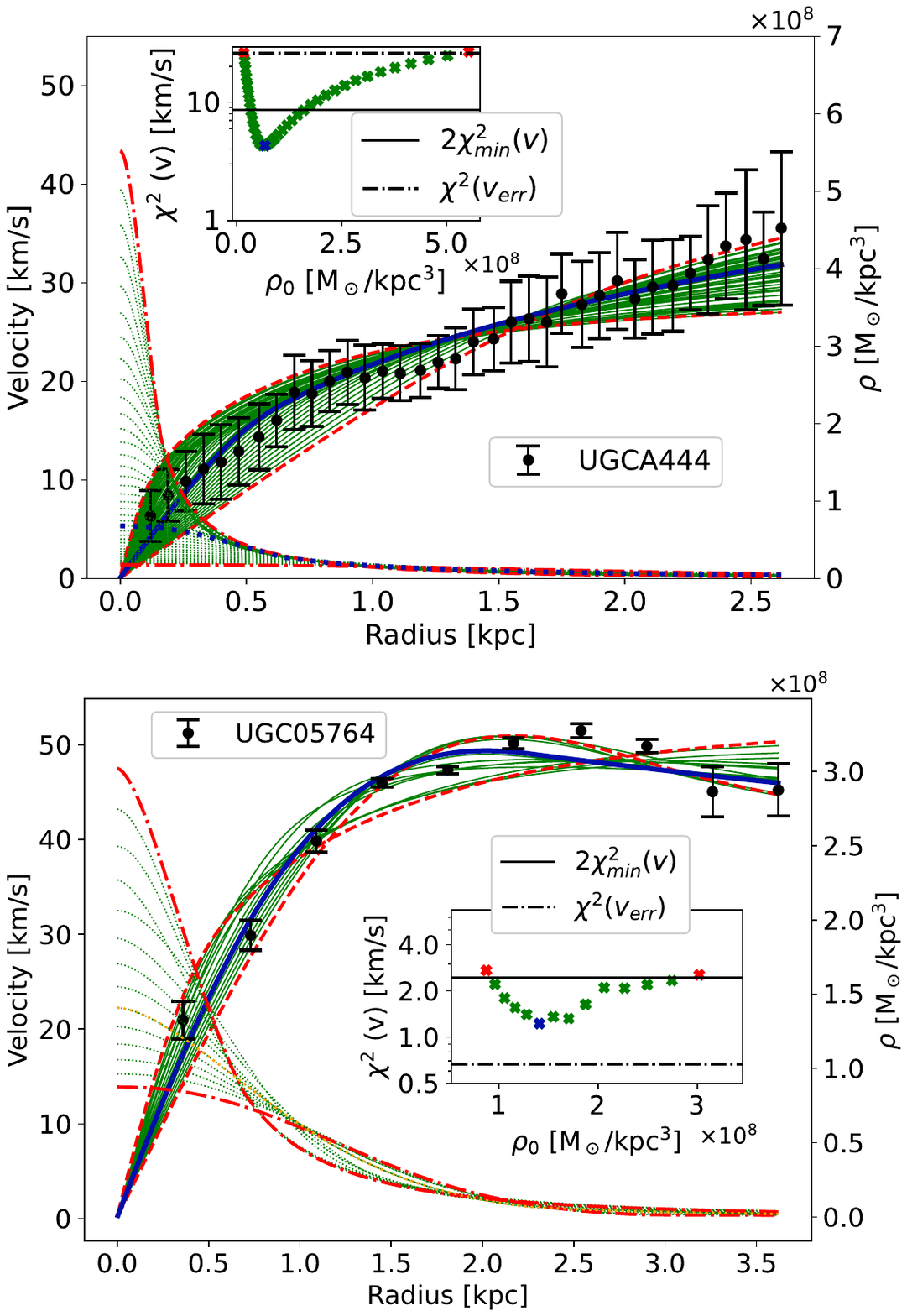}
    \caption{
    Rotation curves (left axis) and corresponding spatial profiles (right axis) for two different galaxies from the SPARC dataset~\cite{SPARC}: 
    optimal (minimum $\chi^2$) fits (blue) are highlighted from a broad list of acceptable fits (green), with the limiting red lines/points denoting incompatibility with our selection rules~\eqref{eq:selection}.
    \vspace{-0.4cm}
    } 
    \label{fig:example}
\end{figure}

\begin{figure*}[t]
    \centering
    \includegraphics[width=0.7\linewidth]{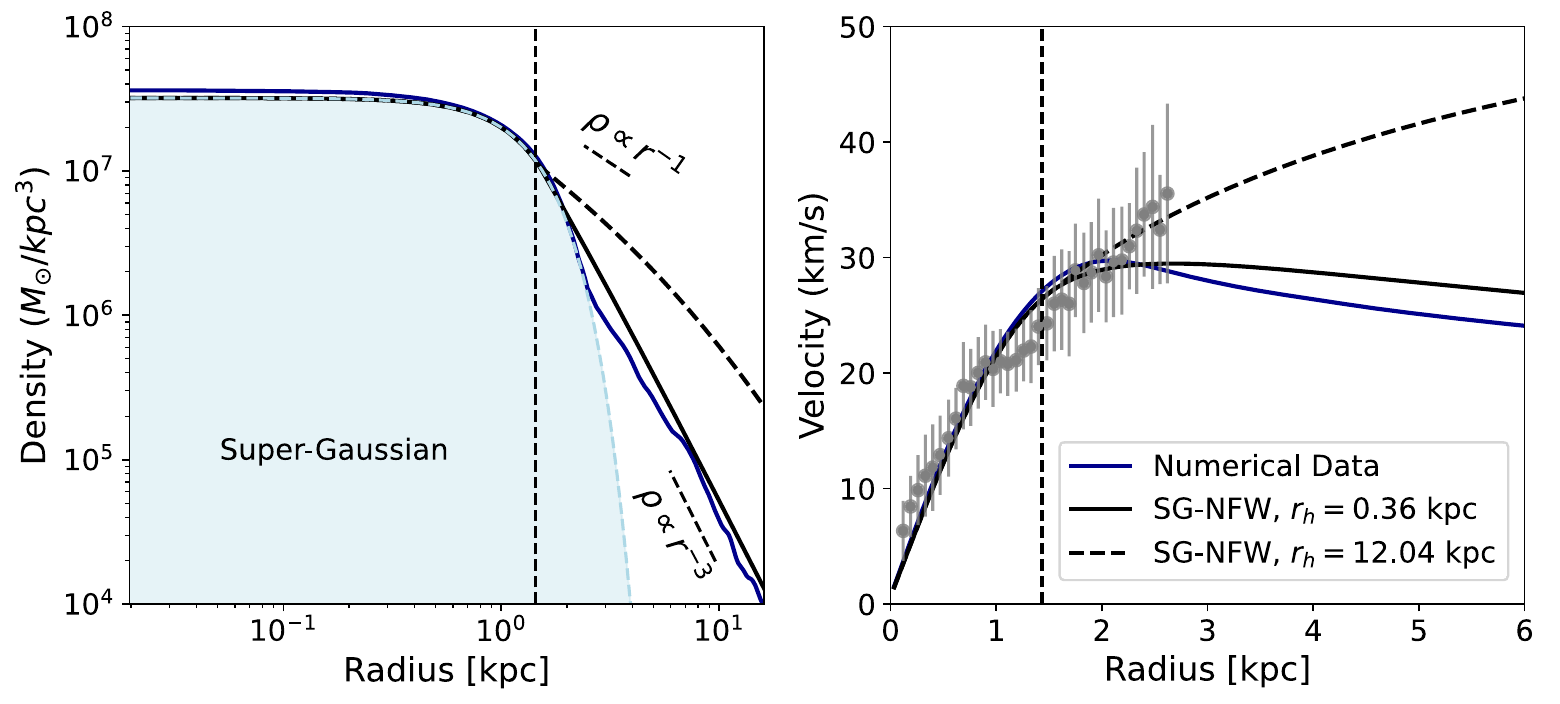}
    \caption{
    Effect of $r_h$ on (a) spatial profiles, and (b) corresponding rotation curves, for the specific example of UGCA444 shown in Fig.~4 of the main text: 
    Shown are static predictions based on SG-NFW profiles for two different values of $r_h$. These correspond to $r_h / r_t \sim 8$ (dashed black lines -- this corresponds to the optimal fit based on $(m_o,g_o)$ shown in Fig.~\ref{fig:all_galaxies_separate}) and to $r_h / r_t \sim 0.2$ (solid black lines -- this corresponds to the target fit of our SFDM simulations shown in Fig.~4(c) of main paper): Such choices clearly show the predominant $r^{-1}$ and $r^{-3}$ scaling of the densities respectively, and thus correspond to a significant difference in the total  halo mass.
    Also shown are the SFDM simulations from Fig.~4(b)-(c) of the main text at a time $\tilde{t} = 0.9$ Gyr  (blue lines), corresponding to the static SG-NFW profiles for the halo mass contained within our numerical grid, based on $r_h / r_t \sim 0.2$. 
Vertical dashed lines in both plots indicate the location of $r_t$, while the SG density profile is shown by the shaded region (and dashed cyan line) in (a).
    }
    \label{fig:rh-variation}
\end{figure*}

\subsection{SG-NFW Fits \& Selection Criteria} 
\label{app: SG-NFW Fits and Selection Criteria}
The allowed SG-NFW parameters, are chosen such that the arising bimodal density plots correspond to rotation curves consistent with observational data, within their error bars. 
Admissible rotation curves are selected via an appropriate fitting tolerance based on~\cite{Freund2003}
\begin{equation}
    \chi^2 (v)= \sum_i \frac{ (v_i - v_{\mathrm{DM},i})^2}{v_{\mathrm{DM},i}} \,,
\end{equation}
where $v_i=v(r_i)$ is the fitting velocity curve from $v(r) = \sqrt{G M(r)\,/\,r}$,
and $v_{\mathrm{DM},i}$ is the dark matter component of the observed velocity curves at each point: each individual plausible numerical rotation curve thus acquires its own $\chi^2(v)$ value, with $v= \sqrt{G M(r)\,/\,r}$ and $M(r)$ fixed by the bimodal SG-NFW profile $\rho(r) = \Theta(r_t-r)\rho_{SG}(r) + \Theta(r-r_t)\rho_{NFW}(r).$

The extent of the observational error bars can be quantified via $\chi^2 (v_{error})$, where $v_{error,i}$ is the maximum velocity allowed by the error bar at point $i$. Fitted velocity curves which differ from the central observed velocity values by a global amount within the observational errors [based on $\chi^2(v)\le \chi^2(v_{error})$] are accepted, leading to the green curves shown for UGCA444 in the main plot of Fig.~\ref{fig:example}(a) [corresponding to Fig.~1(a) of main paper)]. 

However, we note that the above criterion alone would be too stringent for some galactic rotation curves with very small individual error bars for which a good global fit can still be found, such as Fig.~\ref{fig:example}(b) [main paper Fig.~1(b)]. Such global fits can be facilitated by introducing here a further  criterion, accepting global fits for a given galaxy that satisfy  $\chi^2(v) < 2 \chi^2_\text{min}(v)$, where $\chi^2_\text{min}(v)$ is the minimum value obtained for any given curve through the parameter search already described in the main text. 

In order to produce unbiased results for all galaxies, the final combined acceptance criterion used for our rotation curves thus becomes 
\be
\chi^2(v) \leq \text{max}\left[\, \chi^2(v_{error}) \, , \,   2\chi^2_\text{min}(v) \, \right] \;,\label{eq:selection}
\ee
{with such selection criteria consistently implemented across both our heuristic 5-parameter and our physical 3-parameter fits.}

The enforcement of this criterion is shown
{for the heuristic 5-parameter fits}
by green symbols in the insets to Fig.~\ref{fig:example} for two galaxies with relatively large (UGCA444) and small (UGC05764) individual velocity error bars. As evident from the main plots these lead to a broad range of acceptable parameters, corresponding in turn to a huge variation in densities {and thus also to solitonic core masses. Interesting (see also Sec.~\ref{app:fits}), our subsequent physical 3-parameter fits based on fixed ($m_o,\,g_o$) significantly reduce such variation in soliton mass (typically by an order of magnitude).
}

\subsection{Impact of $r_h$ on Rotation Curves}
\label{app: Impact of r_h term}
While our interacting solitonic core region is uniquely determined by the SG profile (defined by $\rho_0$, $r_c$, $\Gamma_g$), fits in the outer halo region are {\em a priori} sensitive to the value of $r_h$, which defines the spatial domain over which the NFW profile transitions from an $\rho(r) \sim r^{-1}$ to $\rho(r) \sim r^{-3}$.
For the majority of the considered galaxies\footnote{However, note that galaxies such as UGC05721 and NGC0024 have a number of data measurements with small error bars at $r \gg r_t$, thus significantly also constraining $r_h$.}, the velocity data are typically restricted within, or just outside, the solitonic core: For such galaxies, we find a broad range of $r_h$ values (spanning almost two orders of magnitude) which are consistent with the observed rotation curves: this {\em a posteriori} justifies our use of $r_h$ as a free fitting parameter (as stated in main text) in trying to identify the optimal parameter sets fitting our selection criteria: ultimately, such selection criteria do constrain $r_h$.

To highlight the role of $r_h$ on the rotation curves, we use the galaxy UGCA444 as a case study.
Fig.~\ref{fig:rh-variation} shows bimodal SG-NFW spatial profiles of our identified SG solitonic core superimposed with the NFW profile for different choices of $r_h$ [Fig.~\ref{fig:rh-variation}(a)], selected such that they produce rotation curves consistent with observational data within their stated error bars [Fig.~\ref{fig:rh-variation}(b)]. 
As clearly evident in this plot, the difference in the NFW density scaling immediately outside the solitonic core (fixed by the ratio of $r_h/r_t$) leads to an observable difference in the corresponding rotation curves in that outer region.

To be more precise, in this figure the black dashed curves highlight the optimal fit through the observational points with $r_h = 12.04$ kpc [see subsequent ideal individual galaxy fit in Fig.~\ref{fig:all_galaxies_separate}], while the solid black curves depict the corresponding prediction with a reduced value $r_h = 0.36$ kpc.
To put this into context, we note that, for this galaxy, $r_t \sim 1.5$ kpc [see subsequent table of optimal individual galactic parameters in Table~\ref{table:parameters}]. In other words, for this galaxy the two different lines correspond to cases $r_h/r_t \sim 8$ (dashed black, optimally traverses midpoints of error bars) to $ r_h/r_t \sim 0.2$ (black, fits within error bars).

In the absence of more accurate observational data from much beyond the solitonic core region (which applies to most SPARC galaxies), there is insufficient information to further constrain $r_h$,
as larger galaxies and higher values of $r_h$ pose significant constraints on required numerical grids for dynamical SFDM simulations {such as those reported here}.
As a result, the proof-of-principle study done in the main manuscript has focused on comparing our dynamical SFDM and static SG-NFW profiles for a smaller value of $r_h$ which still produces rotation curves within the observational errors.

To avoid numerical issues at the simulation boundaries, in this work we have imposed the requirement that  the peak density decreases by four orders of magnitude within the box. 
Choosing the smallest possible $r_h$ value allowed by the rotation curves permits us to achieve this within the smallest possible box. 
We note in this context that accurate dynamical SFDM simulation of the $r_h / r_t \sim 8$ case (dashed black lines in Fig.~\ref{fig:rh-variation}) would require a grid roughly   20 times larger than the one used here.

\begin{figure}
    \centering
    \includegraphics[width=0.95\linewidth]{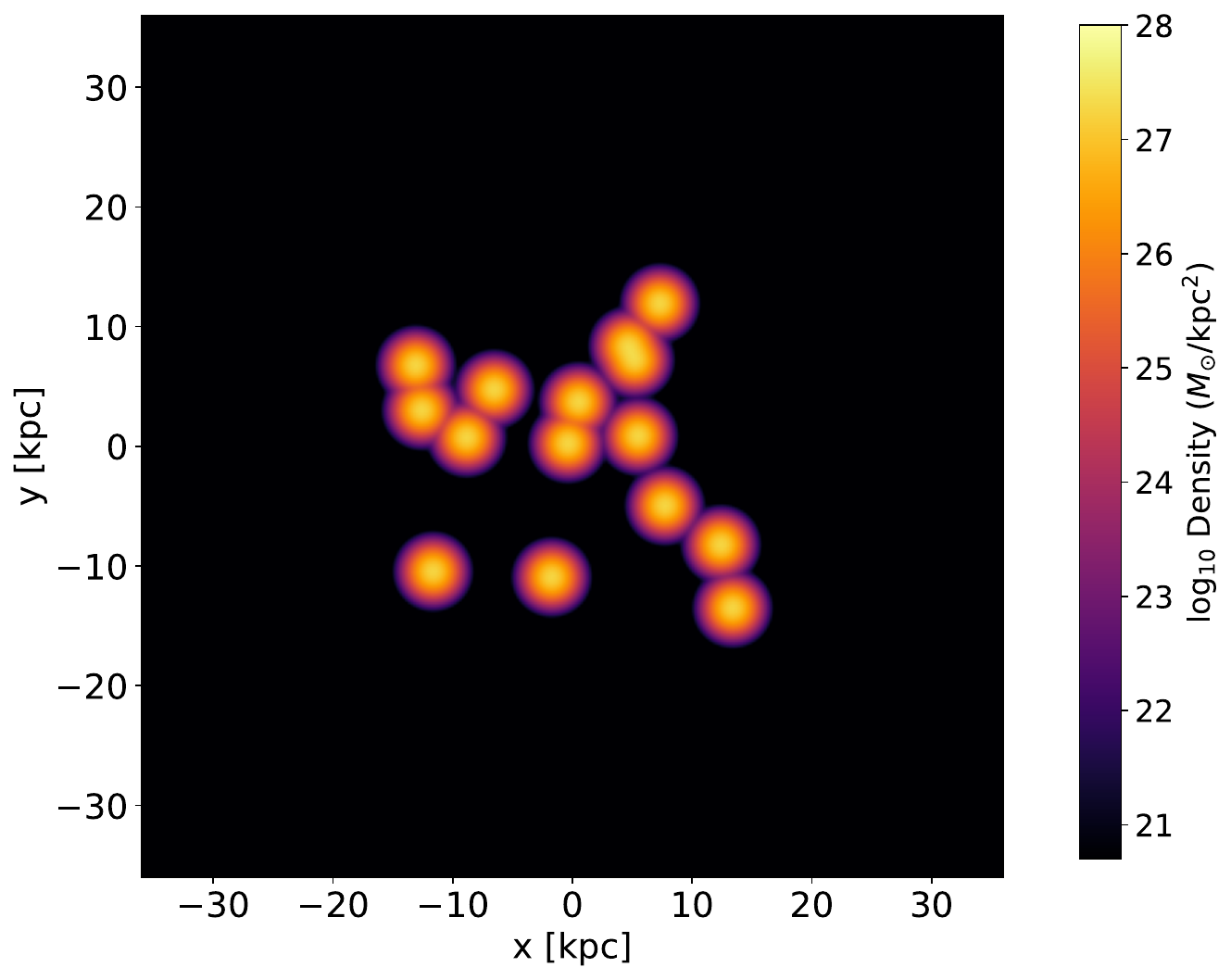}
    \caption{An integrated (along the $z-$axis) projection plot of the initial condition used to generate the density profile shown in Fig.~4(a) of the main paper. The initialization is with fifteen Gaussians distributed randomly such that their center of mass is at the middle of the computational box; the total mass of the simulation is equally distributed between them. }
    \label{fig:initial_ugca444}
\end{figure}

\begin{figure*}[t]
\centering
\includegraphics[width=0.7\linewidth]{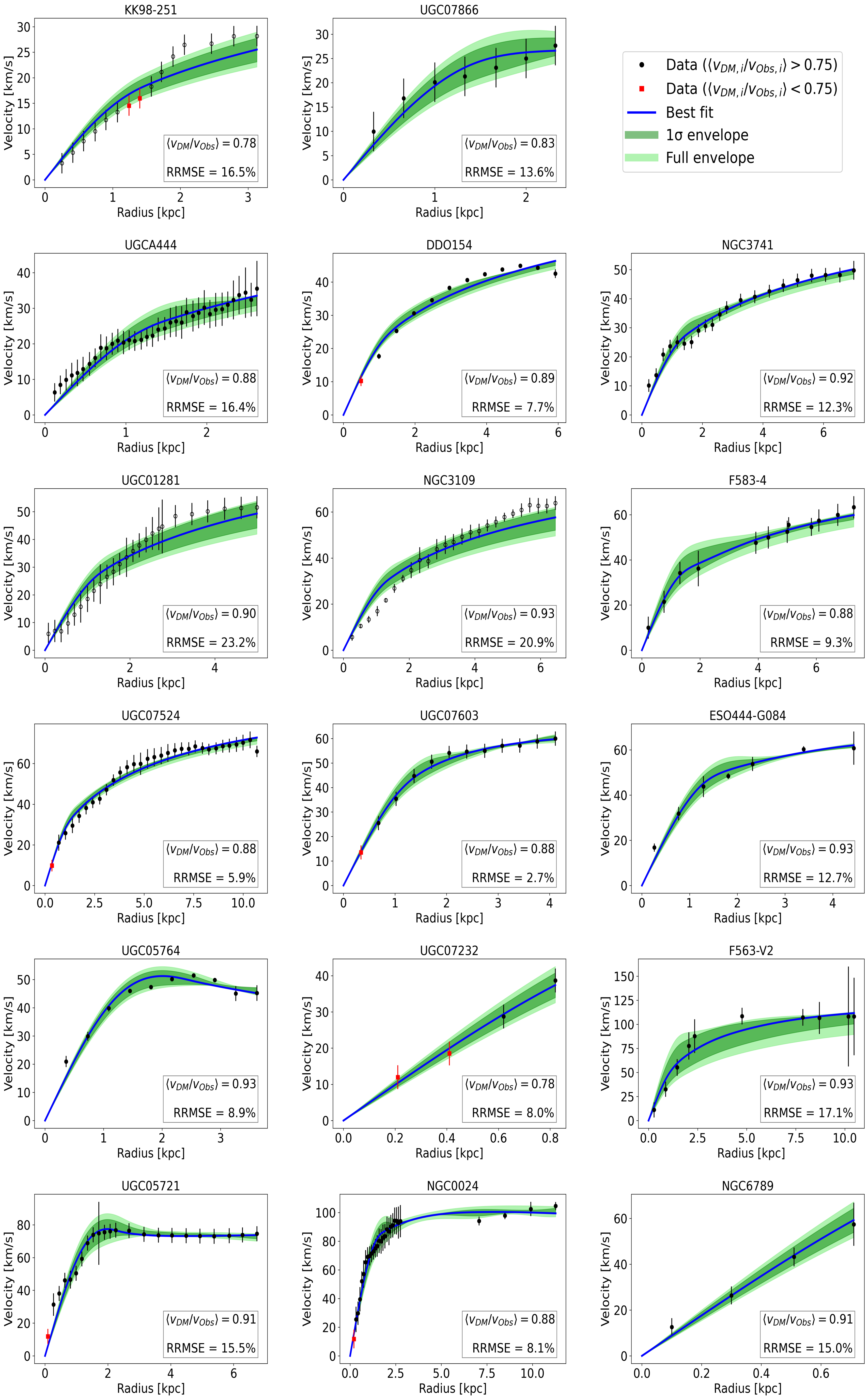}
\caption{
Individual galaxy plots of the optimal {(physical, 3-parameter)} fits to each of the 17 dark-matter-dominated SPARC galaxy rotation curves shown in Fig.~3 of the main paper, based on our SG-NFW static profiles calculated for our extracted optimal values $(m_o,g_o)=\left(1.98 \times10^{-22}\,{\rm eV}/c^2,\, 9.08 \times 10^{-10}\,{\rm eV m}^3/{\rm kg} \right)$.
{All cases reveal very good global fits, with the respective RRMSE and mean value of $v_{\rm DM}/v_{\rm obs}$ over all data for each galaxy shown in the individual plots.
We highlight (with hollow symbols) three of the probed galaxies (KK98-251, UGC01281, NGC3109), which [for this $(m_o,g_o)$ value] reveal a slight overestimation of the velocities deep within the solitonic core, with a corresponding small underestimate outside the core region and slightly higher corresponding RRMSE (in the range (17-23)\%). Individual datapoints which have a ratio of $v_{\rm DM}/v_{\rm obs} < 0.75$ are highlighted in red, with their corresponding values being $v_{\rm DM}/v_{\rm obs} =$ KK98-251: $0.74$ and $0.73$, DDO154: $0.74$, UGC07524: $0.68$, UGC07603: $0.69$, UGC07232: $0.73$ and $0.71$, UGC05721: $0.72$, NGC0024: $0.68$). } 
}
\label{fig:all_galaxies_separate}
\end{figure*}

\subsection{Dynamical GPPE Simulation Details} 
\label{app: Dynamical GPPE Sim.}
The approximate analytical SG-NFW profiles fitting the velocity curves discussed above, can in fact be interpreted as approximations to actual numerical solutions of the full dynamical GPPE equations [Eqs.(1)-(2) in main text]. 
In our proof-of-principle demonstration discussed in the main text, we showed the {possibility of constructing} numerical solutions which are compatible with the rotation curves of the UGCA444 and UGC07866 galaxies {(from long-time evolution starting from soliton merger initial conditions): the choice to probe these specific galaxies} was motivated by their (relatively) low mass within the dark-matter-dominated SPARC galaxy database [see subsequent Table~\ref{table:parameters}], minimizing constraints on numerical grids.

As explained in the main text, although we know the target density profile 
 (and hence also the potential, self-interaction and (soliton) quantum kinetic energy densities),  we do not actually know the total halo mass, nor its kinetic energy. Thus, necessarily, the halo mass of the galaxy included in our numerical simulations is dependent on the spatial extent of our numerical box. 
 Boundary effects are minimised by imposing that the projected density of the SG-NFW, extrapolated to the edge of the box, should be less than  $\mathcal{O}(10^{-4})$ that of the central peak density. 
 For the presented galaxies (and most other galaxies in our dataset), the absence of observed velocities much beyond the solitonic core implies that $r_h$ 
is only {\em a posteriori} constrained  by our selection criteria~\eqref{eq:selection}.
This enables us to minimise grid requirements for the dynamical SFDM simulations, by picking SG-NFW target fits with the smallest value of $r_h$ (i.e.~the smallest mass in the region beyond observational data) which reproduces the desired rotation curve within observational error bars.

Setting possible initial conditions also requires knowledge of the halo's total (conserved) energy, a quantity unknown however since the target profile does not carry information about the halo's internal (classical) kinetic energy. Noting that the soliton constitutes only a small fraction of the total mass 
($M_{c}/M_{\text halo} \ll 1$)
implies that the halos are rather "puffy" i.e. highly excited above the ground state. This requires ${\left| E \right|}/{\left| E_0\right| }\ll 1 $,
where $E<0$ is the energy of the whole halo and $E_0$ is the energy of a hypothetical soliton formed from the whole mass of the halo that constitutes the ground state of the system. In our numerical experiments we find that viable rotation curves can already be obtained for ${\left| E \right|}/{\left| E_0\right| }\sim \mathcal{O}(10^{-2})$.

We can now apply the above approach to numerically obtain potential host halos for the  chosen galaxies through solutions to the GPPE equations. 
We choose the optimal boson mass and self-interaction strength $(m_o,g_o)$ from the analysis of the main text (filled circle in Fig.~2). 
For each galaxy, we extrapolate its fitted density profile out to a radius $L$ at which $\rho(r=0)/\rho(r=L) \approx 10^4$, which uniquely defines the minimum box size ([$-34$,$+34$] ([$-24$,$+24$]) kpc for UGCA444 (UGC07866)), 
with the corresponding mass, $M_{\rm box}$, contained within such a region as an approximate guide for our dynamical simulations; simulations are performed with $N = 320^3$ points and periodic boundary conditions\footnote{The impact of periodic boundary conditions on such simulations has been discussed in~\cite{Rios2023}, with our stated condition $\rho(r=0)/\rho(r=L) \approx 10^4$ necessary to minimize any effect.}.
As initial condition we partition the total mass
into $\mathcal{O}(10)$ equal mass lumps with Gaussian profiles which are randomly distributed within a smaller cube in the box, in such a manner that the centre of mass is kept at the centre of the box. 
An example of an initial condition for our SFDM simulations is shown in Fig.~\ref{fig:initial_ugca444}.
As mentioned in the main text, the energy of the configuration is controlled via the position/number of lumps (modulating the gravitational potential energy) while changing the size of each lump modulates the quantum kinetic and self interaction energies.

\begin{table*}[]
\begin{tabular}{|l||c|c|c|c|c|c|c|}
\hline
            & $M_c$ \; [$10^9 M_\odot]$ & $\rho_0$ \; $[10^8 M_\odot / \text{kpc}^3]$ & $r_c$ \; [kpc]& $r_t$ \; [kpc] & $r_h$ \; [kpc] & \; \; $\vartheta$ \; \; & \; \; \; $\Gamma_g$ \; \; \;\\ \hline \hline
KK98-251    & 0.27                                                                     & 0.15                           & 1.21  & 1.47  & 12.10           & 1.97        & 4.8        \\ \hline
UGC07866    & 0.45                                                                     & 0.28                           & 1.20  & 1.42  & 0.30            & 2.06        & 12.9       \\ \hline
UGCA444     & 0.51                                                                     & 0.32                           & 1.20  & 1.50  & 12.04           & 2.08        & 16.6       \\ \hline
DDO154      & 0.53                                                                     & 0.34                           & 1.20  & 1.41  & 12.05           & 2.09        & 18.0       \\ \hline
NGC3741     & 0.55                                                                     & 0.35                           & 1.20  & 1.40  & 12.05           & 2.09        & 19.6       \\ \hline
UGC01281    & 0.62                                                                     & 0.40                           & 1.21  & 1.40  & 12.06           & 2.11        & 25.2       \\ \hline
NGC3109     & 0.74                                                                     & 0.48                           & 1.21  & 1.39  & 12.07           & 2.14        & 35.2       \\ \hline
F583-4      & 0.84                                                                     & 0.55                           & 1.21  & 1.53  & 12.08           & 2.15        & 45.2       \\ \hline
UGC07524    & 0.90                                                                     & 0.60                           & 1.21  & 1.38  & 12.09           & 2.16        & 53.4       \\ \hline
UGC07603    & 1.44                                                                     & 0.97                           & 1.21  & 1.37  & 2.67            & 2.21        & 133.6      \\ \hline
ESO444-G084 & 1.56                                                                     & 1.05                           & 1.21  & 1.72  & 5.03            & 2.21        & 157.8      \\ \hline
UGC05764    & 1.56                                                                     & 1.05                           & 1.21  & 2.56  & 0.30            & 2.21        & 157.8      \\ \hline
UGC07232    & 2.01                                                                     & 1.36                           & 1.21  & 1.36  & 0.30            & 2.23        & 260.3      \\ \hline
F563-V2     & 2.34                                                                     & 1.62                           & 1.22  & 1.36  & 9.78            & 2.24        & 363.4      \\ \hline
UGC05721    & 3.45                                                                     & 2.37                           & 1.22  & 2.25  & 7.42            & 2.25        & 769.9      \\ \hline
NGC0024     & 4.25                                                                     & 2.92                           & 1.22  & 1.61  & 2.68            & 2.26        & 1168.2     \\ \hline
NGC6789     & 5.03                                                                     & 3.46                           & 1.22  & 1.35  & 0.30            & 2.26        & 1630.8       \\ \hline
\end{tabular}
\caption{
Table of extracted parameter values for each of the 17 SPARC galaxies based on our SG-NFW fits. These parameters allow the full reconstruction of static spatial profiles, and rotation curves, highlighting at the same time their support for SFDM through $\Gamma_g \gg 1$.
Note that while the quoted values of $r_h$ correspond to ones used to obtain the optimal rotation curves shown in Fig.~\ref{fig:all_galaxies_separate} based on a minimum $\chi^2(v)$ fit criterion, the small number of points at values $r \gg r_t$ (only available for some galaxies) combined (in some cases) with sizeable error bars in the observational velocities, imply that the rotation curves can, for many of those galaxies, still be well reproduced within the observational error bars by values of $r_h$ differing by at least an order of magnitude from the ones quoted in this table. An example of this is shown, for UGCA444, in Fig.~\ref{fig:rh-variation}.}
\label{table:parameters}
\end{table*}

We initialise the SFDM simulation with an appropriate variety of initial conditions (allowing for slight variations in the total mass and total energy) and propagate dynamically for an extended time of 
$ {\cal O} (10) \, \text{Gyr}$, in order to ensure that the emerging quasi-equilibrium core-halo structure arising from the gravitational coalescence of the initial masses~\cite{Liu2023} leads to the desired parameters.
At that point,
we begin sampling the shell averaged density profile from consecutive timesteps, and compare the computed velocity curve to the observational data, and thus identify a temporal regime of ${\cal O}(1)$ Gyr over which the rotation curves remain consistent with observational data (within their respective error bars): final results have been shown in the main text (Fig.~4) [see also corresponding comparison of solid blue and black curves in Fig.~\ref{fig:rh-variation}.]
We highlight here that while the peak density grows gradually as the system undergoes dynamic cooling: this occurs over timescales larger than the observationally relevant ones, as suggested in~\cite{Levkov2018, Eggemeier2019, Chen2021, Rios2024}.

\subsection{Optimal Extracted Parameters \& Rotation Curve Fits for Individual SPARC Galaxies} \label{app:fits}

In Fig.~\ref{fig:all_galaxies_separate} we plot for each galaxy individually the optimal rotation curve fits already shown {collectively} in Fig.3 of the main text, based on our identified $(m_o,g_o)$ value. The corresponding parameters for their SG cores allowing their static profile and rotation curve reconstruction are given in Table.~\ref{table:parameters}. {The average ratio of $v_{\rm DM}/v_{\rm obs}$ for each galaxy is shown in each subplot, with  $0.78 \le \langle v_{\rm DM}/v_{\rm obs} \rangle \le 0.93$, with 15 of the 17 galaxies having $\langle v_{\rm DM}/v_{\rm obs} \rangle \ge 0.88$.
Additionally, in this figure, we highlight those few {\em individual} data points with $v_{\rm DM}/v_{\rm obs} < 0.75$, noting that these correspond to (at most) 1-2 points per galaxy (all with $v_{\rm DM}/v_{\rm obs} \ge 0.68$), with such points -- located within their respective solitonic cores -- not noticeably affecting the global rotation curves.

The RRMSE \eqref{eq:rrmse} 
of the individual galaxy best fit curves 
is given in table \ref{table:parameters2} leading to an average value of $\left\langle{\rm D}[v_{\rm bf}]\right\rangle =0.13$ and standard deviation $\left \langle {\rm D}^2[v_{\rm bf}]  - \left\langle{\rm D}[v_{\rm bf}]\right\rangle^2\right\rangle^{1/2} = 0.05$. It is worth comparing ${\rm D}[v_{\rm bf}]$ (obtained from our physical 3-parameter fit with fixed $(m_o,g_o)$) with the corresponding value, $D_{5}[v_{\rm bf}]$, resulting from the original heuristic 5-parameter fits of Sec.~\ref{sec:III} in which $m$ and $g$ were allowed to vary per galaxy. As expected, given the larger number of available parameters, the 5-parameter fits are slightly closer to the observed rotation curves, with $D_{5}[v_{\rm bf}]\simeq (0.07 \pm 0.04)$ -- see Table~\ref{table:parameters2}. However, as stressed in the text, varying $m$ and $g$ is not physical. Interestingly, the physical model with fixed $m$ and $g$ allows for a significantly narrower confidence interval for the core mass $M_c$ (by an order of magnitude) - see Table~\ref{table:parameters2}. 
} 

\newpage
\begin{table*}[b]
\centering
\renewcommand{\arraystretch}{1.2}
\begin{tabular}{|l||c|c|c|c|}
\hline
\textbf{Galaxy} & \textbf{$D(v_{\mathrm{bf}})$} & \textbf{$D_5(v_{\mathrm{bf}})$} & \textbf{$\Delta \log_{10} M_c$ [M$_\odot$]} & \textbf{$\Delta_5 \log_{10} M_c$ [M$_\odot$]}  \\ 
\hline\hline
KK98-251      & 0.1651 & 0.0498 & 0.1450 & 1.5525 \\
UGC07866      & 0.1355 & 0.0222 & 0.2138 & 2.2027 \\
UGCA444       & 0.1639 & 0.0955 & 0.1791 & 1.8768 \\
DDO154        & 0.0765 & 0.0340 & 0.0924 & 0.2961 \\
NGC3741       & 0.1225 & 0.0870 & 0.1481 & 0.7326 \\
UGC01281      & 0.2320 & 0.1670 & 0.1718 & 1.3349 \\
NGC3109       & 0.2095 & 0.0487 & 0.1638 & 0.6590 \\
F583-4        & 0.0933 & 0.0724 & 0.2722 & 1.5157 \\
UGC07524      & 0.0594 & 0.0593 & 0.0876 & 1.1694 \\
UGC07603      & 0.0275 & 0.0205 & 0.1218 & 0.6554 \\
ESO444-G084   & 0.1272 & 0.0348 & 0.1438 & 1.0924 \\
UGC05764      & 0.0894 & 0.0647 & 0.1077 & 0.5239 \\
UGC07232      & 0.0803 & 0.0789 & 0.1614 & 5.4956 \\
F563-V2       & 0.1705 & 0.0291 & 0.4412 & 1.8995 \\
UGC05721      & 0.1547 & 0.0815 & 0.1718 & 1.6808 \\
NGC0024       & 0.0807 & 0.0857 & 0.1744 & 0.9610 \\
NGC6789       & 0.1498 & 0.1153 & 0.1638 & 6.0049 \\
\hline
\textbf{Mean}     & 0.1258 & 0.0674 & 0.1741 & 1.7443 \\
\textbf{Std. Dev. }  & 0.0530 & 0.0365 & 0.0792 & 1.5545 \\
\hline
\end{tabular}
\caption{{The RRMSE \eqref{eq:rrmse} for the best fit curve of our physical 3-parameter model ($D(v_{\mathrm{bf}})$) and a confidence region $(\Delta \log_{10} M_c)$ for the central soliton mass $M_c$, computed as the FWHM of the distribution of the $\chi^2(M_c)$ values for the envelope of acceptable rotation curves. For comparison, we also show the RRMSE for the best fit curve of the non-physical 5-parameter fits ($D_5(v_{\mathrm{bf}})$), as well as their confidence region $(\Delta_5 \log_{10} M_c)$ for $M_c$. The lower two entries of each column show the corresponding mean and and standard deviation for the 17 galaxies.} }
\label{table:parameters2}
\end{table*}

\providecommand{\newblock}{}


\begin{thebibliography}{10}
\bibitem{footnote_probability}
Note that this should not to be strictly interpreted as a probability distribution, as we are not necessarily sampling the underlying space uniformly with our constructed degeneracy curves.
\end{thebibliography}
\end{document}